\RequirePackage{fix-cm}
\documentclass[twocolumn,epjc3]{svjour3}
\makeatletter
\expandafter\let\csname opt@amsmath.sty\endcsname\relax
\AtBeginDocument{
  \mathindent=15pt 
  \@mathmargin\@centering} 
\makeatother

\smartqed  
\RequirePackage{graphicx}
\usepackage{tikz}
\usetikzlibrary{patterns}
\usepackage{axodraw}
\usepackage{cite}
\usepackage{epsfig,multicol}
\usepackage{blindtext}
\usepackage{lineno}
\usepackage{amsmath,amssymb,mathrsfs}

\newcommand\lsim{\mathrel{\raise.3ex\hbox{$<$\kern-.75em\lower1ex\hbox{$\sim$}}}}
\newcommand\gsim{\mathrel{\raise.3ex\hbox{$>$\kern-.75em\lower1ex\hbox{$\sim$}}}}

\newenvironment{Eqnarray}%
     {\arraycolsep 0.14em\begin{eqnarray}}{\end{eqnarray}}
\newcommand{\beqa}{\begin{Eqnarray}}
\newcommand{\eeqa}{\end{Eqnarray}}
\newcommand{\beq}{\begin{equation}}
\newcommand{\eeq}{\end{equation}}

\journalname{Eur. Phys. J. C}
\begin{document}
\title{
High $P_T$ Higgs excess as a signal of non-local QFT at the LHC
%
}
%
\titlerunning {High $P_T$ Higgs excess as a signal of non-local QFT at the LHC}
\author{Xing-Fu Su\thanksref{addr1}
  \and
  You-Ying Li\thanksref{addr1}
  \and
  Rosy Nicolaidou\thanksref{addr2}
 \and
  Min Chen\thanksref{addr1}
 \and
  Hsin-Yeh Wu\thanksref{addr1}
  \and
  Stathes Paganis\thanksref{e3,addr1}
}
%
\thankstext{e3}{stathes.paganis@cern.ch, corresponding author}
\authorrunning{You-Ying Li, R.Nicolaidou, S.Paganis, Hsin-Fu Su, Min Chen}
\institute{
 Department of Physics, National Taiwan University, 
No 1, Sec 4, Roosevelt Road, Taipei 10617, Taiwan\label{addr1}
 \and ~IRFU, CEA, Universite Paris-Saclay, Gif-sur-Yvette; France\label{addr2}
}
\date{Received: date / Accepted: date}

\maketitle

\begin{abstract}
Non-local extensions of the Standard Model with a non-locality 
scale $\Lambda_{NL}$ have the         
effect of smearing the pointlike vertices of the Standard Model.
At energies significantly lower than $\Lambda_{NL}$ vertices appear
pointlike, while beyond this scale all beta functions
vanish and all couplings approach a fixed point leading to
scale invariance. Non-local SM extensions are ghost free, with the
non-locality scale serving as an effective cutoff to radiative
corrections of the Higgs mass. 
We argue that the data expected to be collected at
the LHC phase 2 will have a sensitivity to non-local 
effects originating from a non-locality scale of a few TeV.
Using an infinite derivative prescription, we study modifications
to heavy vector-boson cross sections that can lead to an
enhanced production of boosted Higgs bosons in a region of the 
kinematic phase space where the SM background is very small. 
\end{abstract} 
%

\section{Introduction}
\label{intro}
Continuing searches at the Large Hadron Collider (LHC) have shown no 
evidence of new physics Beyond the Standard Model (BSM). The success
of the SM suggests that the scale of new physics $\Lambda_{NP}$ must
be high enough when compared to the electroweak scale $\Lambda_{EW}$,
so that its effect on the observed data be small. Modifications
of SM cross sections at high momentum transfers (e.g. modifications 
of Higgs and $W/Z$ boson yields), 
could be the first sign of new physics and new heavy particles
\cite{HbbBroadRes}, \cite{OurCHpaper1}.


A generic extension of the SM in which the particle interaction
vertices are smeared is not a new idea \cite{Biswas}. 
A plethora of new physics scenaria involving new particles and new 
interactions can fit in such an effective description. To give a few examples: 
GUT theories \cite{Pati,Georgi,Fritzcsh}, composite Higgs
\cite{Eichten,Contino}, little Higgs \cite{lHiggs1,lHiggs2,lHiggs3}, and models
with vector $Z^\prime$ \cite{Zprime1, Zprime2}, and $W^\prime$ \cite{Wprime}.
Independent of the new physics behind this smearing, 
one can modify the pointlike behaviour by introducing higher
derivatives in the kinetic terms of the SM Lagrangian 
or in the interaction vertices \cite{Buoni1}.
The simplest example is the smearing of a pointlike source 
represented by the Dirac delta function, by the exponential of an
entire analytic function of derivatives:
\begin{equation}
\nonumber
e^{\alpha^2 \partial_x^2}\delta(x) 
=
e^{\frac{\partial_x^2}{\Lambda^2}}\delta(x) 
= 
\frac{1}{\alpha\sqrt{2}} e^{-\frac{x^2}{4\alpha^2}}.
\end{equation}
Application of this infinite sum of derivatives on the delta
function, 
leads to the smearing of a pointlike to a Gaussian source. The energy scale 
at which smearing effects become important is $\Lambda = 1/\alpha$.
Introduction of higher derivative terms in the Lagrangian leads to 
non-local effects that become relevant at a scale $\Lambda \simeq \Lambda_{NL}$.
Extending the traditional QFT to a non-local version 
with a non-locality scale $\Lambda_{NL}$ has several attractive
features \cite{Buoni1}.
There are no new dynamical degrees of freedom 
other than the original ones of the corresponding local QFT, and the
theory is UV finite. Beyond the non-locality scale all beta functions 
vanish, and all couplings approach a fixed point determined by  
$\Lambda_{NL}$ \cite{GhoshalNA, GhoshalA}.

In practice, the non-local modification of the Lagrangian is achieved by
introducing an infinite series of derivatives in the kinetic terms of 
the form:
\begin{equation}
e^{f(D^2)} = e^{\pm \frac{D^2}{\Lambda_{NL}^2}},
\label{NLfunction}
\end{equation}
where $D_{\mu}=\partial_{\mu}-igT^a A_{\mu}^a$ is the covariant
derivative and $D^2 = \eta_{\mu\nu}D^{\mu}D^{\nu}$.
This particular choice  $e^{f(D^2)}$ where $f(D^2)$ is an entire
analytic function, is required for avoiding the appearance of 
any extra poles in the propagator \cite{Buoni1}.
The fermionic part of the Lagrangian is modified as follows:
\begin{equation}
\mathcal{L} = i\bar{\psi} e^{\frac{D^2}{\Lambda^{2}_{NL}}}\gamma^{\mu}D_{\mu}\psi
\end{equation}
In the limit $\Lambda_{NL}\rightarrow \infty$ the original 
Lagrangian, in our case the SM, is recovered.

If we proceed to extend the SM Lagrangian to a non-local version (NL-SM)
by introducing an infinite derivative sum in the kinetic terms, 
shown in Eq.~\ref{NLfunction},
the result is a smearing of interaction vertices affecting 
relevant measured cross sections.
The Drell-Yan $ff\rightarrow \ell\ell$ process can set limits to the
non-locality scale, in particular dilepton measurements at the LHC 
for dilepton invariant mass 
$M_{\ell\ell}>1$~TeV \cite{ATLAS-dilepton}, \cite{ATLAS-dilepton2}, \cite{CMS-dilepton}. 
At the very high energies offered at the LHC, the weak gauge-boson
scattering $VV\rightarrow VV$ is dominated by the longitudinal degrees of
freedom. So, effectively this scattering is Higgs-Higgs field
scattering. In these processes the $VVV$, $VVH$ vertices may 
not be pointlike as a consequence of the new BSM physics.
It should be stressed that non-locality in this context does not 
mean that the spacetime is discontinuous. Here we still consider 
a continuous spacetime, while non-locality is introduced through 
the presence of form factors due to new interactions appearing 
at a TeV or multi-TeV scale.
 
While approaching the non-locality scale from a low\-er energy, 
the interaction vertices appear smeared, and the theoretical
treatment/modeling is typically done using form factors. This picture
is similar to low energy electron-proton scattering where as the
$Q^2$ increases, the non-pointlike nature of the proton is
revealed. Non-local modification approaches have also been used in nuclear 
reactions to explain proton-deuteron scattering data \cite{nuclearpd}.
Theoretical models that can lead to such form factors have
been proposed in the literature and many are gravity or string
motivated~\cite{Witten}, \cite{Eliezer}. 
In string field theory $a^\prime = 1/\Lambda^2_{NL}$, is the universal 
Regge slope. In the case of gravity, non-locality may appear at a Grand Unification
scale at energies close to the Planck scale, while in the case of the
SM, non-locality in the TeV scale is due to new particles and new 
interactions present at this scale. 
Thus one still needs to explain the origin of such non-locality, with 
compositeness being a good candidate.

New heavy vector bosons with masses close to the new physics scale
$\Lambda_{NL}$ appear in many BSM scenaria. 
In composite models they are expected to be present as the $J=1$ poles of a 
$\rho$-like Regge trajectory.
In this work, we propose to study the effects of non-local QFT 
modification on heavy vector-boson triplets (HVT). 
The impact of smeared vertices on these heavy states is to change their production
cross section and their mass lineshape. In non-local QFT theories as
in \cite{Biswas}, \cite{Buoni1},  the modification enters through an
exponential factor
of the usual form: $e^{\pm D^2/\Lambda_{NL}^2}$, where the scale of
new physics $\Lambda_{NL}$ can be approximated by the pole mass of the heavy vector
boson. Depending on the sign of the exponential and the mass of the
heavy vector boson with respect to the non-locality scale, 
this factor has the
effect to increase or decrease the cross-section predicted by the
local theory and change the lineshape of the resonance. 
The actual effect depends on the details of the new
physics and only measurements of these modifications can provide
more information about the new theory.

In this paper, we first introduce the non-local QFT approach and 
summarize the potential impact of recent LHC results. In Section~3 we present
an analysis used to search for heavy vector triplets at the LHC, and we
report on the search potential for anomalous boosted Higgs production 
in association with weak gauge bosons 
as a function of the luminosity and the non-locality scale $\Lambda_{NL}$.
Finally we conclude in Section~4.

\section{Signals of Non-locality and Constraints from Data}
\label{hvt}

\begin{figure}[hbt!]
\begin{center}
\resizebox{0.53\textwidth}{!}{
\includegraphics{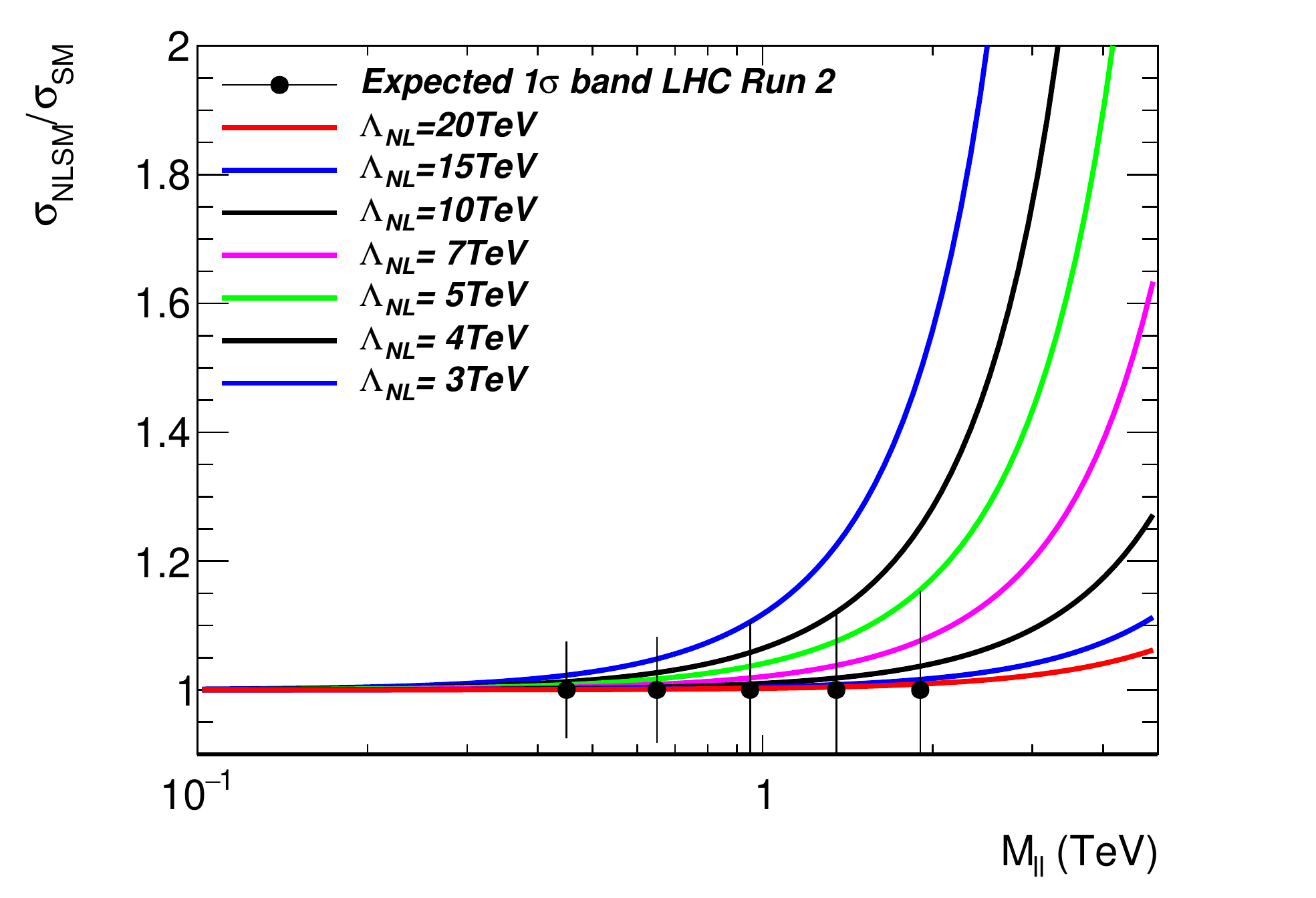}
}
\caption{Drell-Yan dilepton cross-section modification with respect to
  the SM prediction due to
  non-local effects for an integrated luminosity of 137~fb$^{-1}$.
The expected statistical and systematic uncertainties from a
measurement at the LHC are also shown~\cite{CMS-dilepton}. 
Based on these expectations, 
present and future measurements
at the LHC can place constraints in the non-locality scale $\Lambda_{NL}$.
}
\label{DYmods}
\end{center}
\vspace{-5mm}
\end{figure}

Non-local modification of the SM leads to modifications of SM cross
sections. The Drell-Yan dilepton cross section may receive positive 
corrections of the following general form \cite{Biswas}:
\begin{equation}
\sigma_{NL-SM} = e^{a (s/\Lambda_{NL}^2)} \times \sigma_{SM},
\label{nlmod}
\end{equation}
where $s$ is the square of the dilepton invariant mass, and 
$a$ is a real factor that generalizes the $\pm 1$ factor of 
Eq.~\ref{NLfunction}. 
Using the full datasets from the Run 2 of LHC, one can already
place upper limits to a hypothetical non-locality scale.
Dilepton data from ATLAS and CMS extend out to an
invariant mass of about 2.2 TeV \cite{ATLAS-dilepton}, 
\cite{ATLAS-dilepton2}, \cite{CMS-dilepton}. 
%
In this work we assume positive $a>0$ and highlight 
the potential of the DY process in constraining the non-locality scale 
$\Lambda_{NL}$. It should be stressed that 
the actual extraction of limits or
excesses using data is left to the LHC experiments.
Using Eq.~\ref{nlmod}, predicted Drell-Yan modifications
are presented in Fig.~\ref{DYmods} for $a=1$ and varying values of the
non-locality scale $\Lambda_{NL}$.
In Fig.~\ref{DYmods} the expected data 1-$\sigma$ combined uncertainty 
for an integrated luminosity of 137~fb$^{-1}$
is shown, based on the yields in the same $M_{\ell\ell}$ bins presented in
\cite{CMS-dilepton}. 
For dilepton masses above 2~TeV the data statistical uncertainty 
becomes dominant.
Examining Fig.~\ref{DYmods}, we see that a single LHC experiment can 
already place significant constraints on the
non-locality scale $\Lambda_{NL}$ at the level of a few TeV.
The current level of deviation of the data from expectation is left 
to the experiments to determine.

As discussed in the introduction, non-locality can cause distortions
to theoretically motivated broad heavy vector resonances. Broad resonances or a
continuum of states are particularly interesting because their decays to
dileptons are suppressed and the main decay mode is the diboson $VH$
and $VV$ \cite{HbbBroadRes}. In particular, the $VH$ channel offers a
unique signal of a very high $P_T$ Higgs produced in association with 
a weak gauge boson, with very low SM background
expectation. It is thus worthwhile to explore the potential of such a
search and develop a method to produce heavy vector bosons with their
lineshapes and cross sections modified by the non-locality scale.
A straightforward approach is to use the HVT model, \cite{hvt1}, that is also
employed by the LHC experiments and then modify the predicted cross
sections using Eq.~\ref{nlmod}. The HVT model 
is briefly summarized below.

HVT is a general phenomenological Lagrangian that can be used for the modeling
of resonances predicted by a wide range of BSM scenarios.
The Lagrangian describing the interactions of these resonances $V^{a \prime}$, $a=1,2,3$ with quarks, 
leptons, vector bosons and the Higgs boson is shown below:
\begin{eqnarray}
\nonumber
\mathcal{L}_{V}^{int}=&-&\frac{g^2c_F}{g_V}V_{\mu}^{a \prime}\bar{q}_k\gamma^{\mu}\frac{\sigma_{a}}{2}q_{k}
-\frac{g^2c_F}{g_V}V_{\mu}^{a \prime}\bar{\ell}_k\gamma^{\mu}\frac{\sigma_{a}}{2}\ell_{k}\\\nonumber
&-&g_Vc_H\left(V_{\mu}^{a \prime}H^{\dagger}\frac{\sigma^a}{2}iD^{\mu}H+\mathrm{hc}\right),\nonumber
\end{eqnarray}
where $q_k$ and $\ell_k$ are the quark and lepton weak doublets, $H$ is the Higgs 
doublet and $\sigma^a$ the three Pauli matrices.
In this Lagrangian, the HVT triplet
$V^{a \prime}=(W^{+\prime},W^{-\prime},Z^{\prime})$ interacts with the Higgs
doublet, i.e. the longitudinal degrees of freedom of the SM $W$ and
$Z$ bosons and the SM Higgs, with a coupling strength $g_V$. In
order to allow for a broader class of models, this coupling strength
can be varied by the parameter $c_H$, so in the Lagrangian the
full coupling to the SM weak and Higgs bosons is $g_Vc_H$.
The HVT resonances also couple to the SM fermions, again
through their coupling to the SM weak and Higgs bosons,
$g^2/g_V$, where $g$ is the SM $SU(2)_L$ weak gauge
coupling. This coupling between HVT resonances and fermions is also
controlled by an additional parameter $c_F$ to allow for a broader
range of models to be included, as follows: $g^2c_F/g_V$.
In this work, we use the so-called Drell-Yan scenario B, also used by 
the experiments \cite{ATLASHeavyCombination}.
This is a strongly
coupled scenario as in composite Higgs models with $\frac{g^2c_F}{g_V}=0.14$ and  $g_Vc_H=-2.9$. 
where the $V^{\prime}$ resonances are broader than 
in weakly coupled scenarios.
For $|g_Vc_H|>3$, the resonance intrinsic 
width becomes significant and cannot be neglected. For such large
couplings, the HVT resonance $VH$ decay mode dominates while the
di-fermion mode is heavily suppressed.
The phenomenological implications of $VH$ mode decay dominance 
have been discussed in \cite{HbbBroadRes}. 

\begin{figure}[hbt!]
\begin{center}
\resizebox{0.5\textwidth}{!}{
\includegraphics{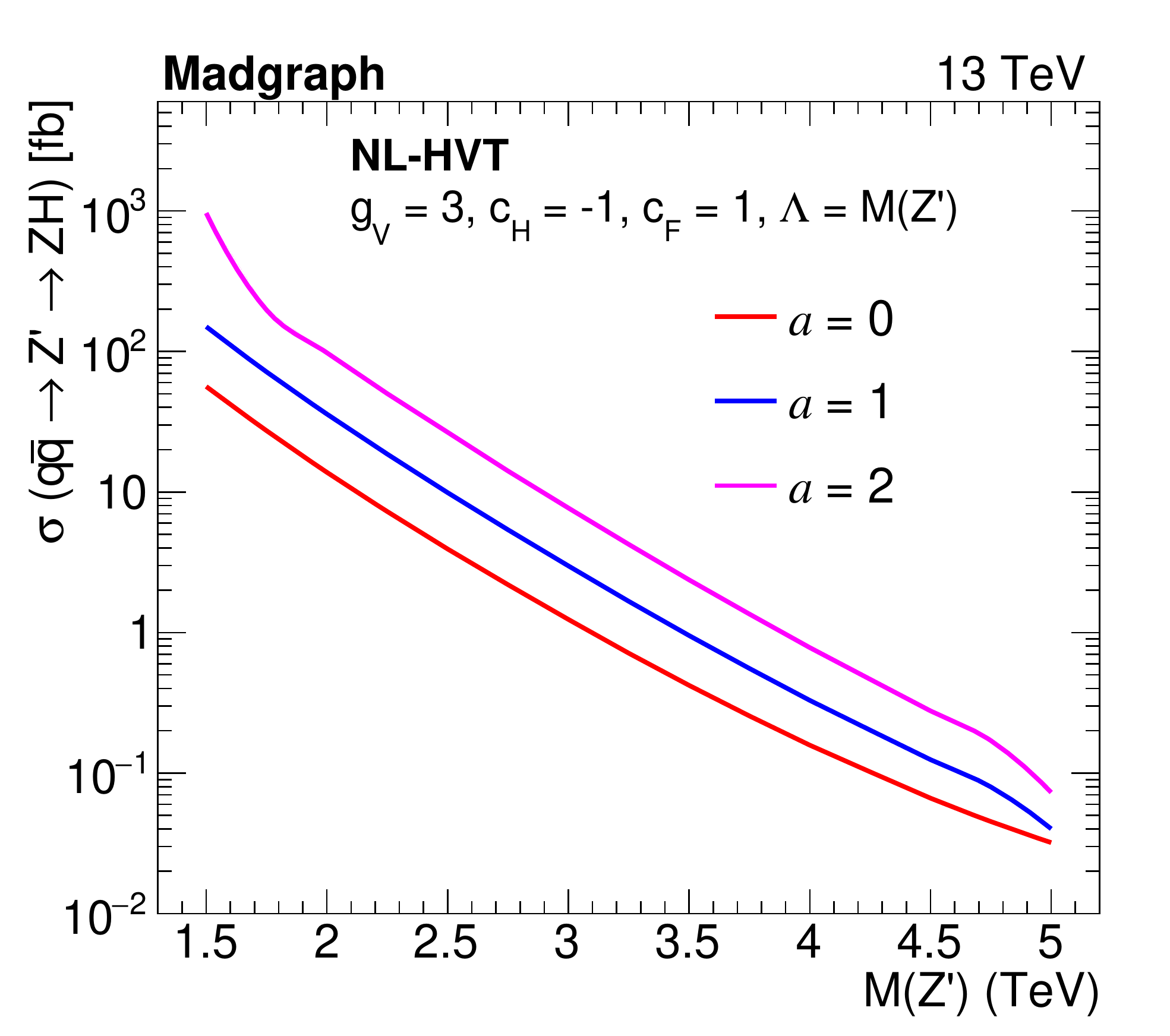}
}
\caption{$Z^\prime$ resonance cross sections for the DY production
  mode after a non-local modification, as a function of the
pole mass $M(Z^\prime)$. The non-locality scale $\Lambda_{NL}$ in
  Eq.\ref{nlmod} has been 
set at the pole mass $M(Z^\prime)$.
}
\label{NLxsection}
\end{center}
\vspace{-5mm}
\end{figure}
\begin{figure}[hbt!]
\begin{center}
\resizebox{0.5\textwidth}{!}{
\includegraphics{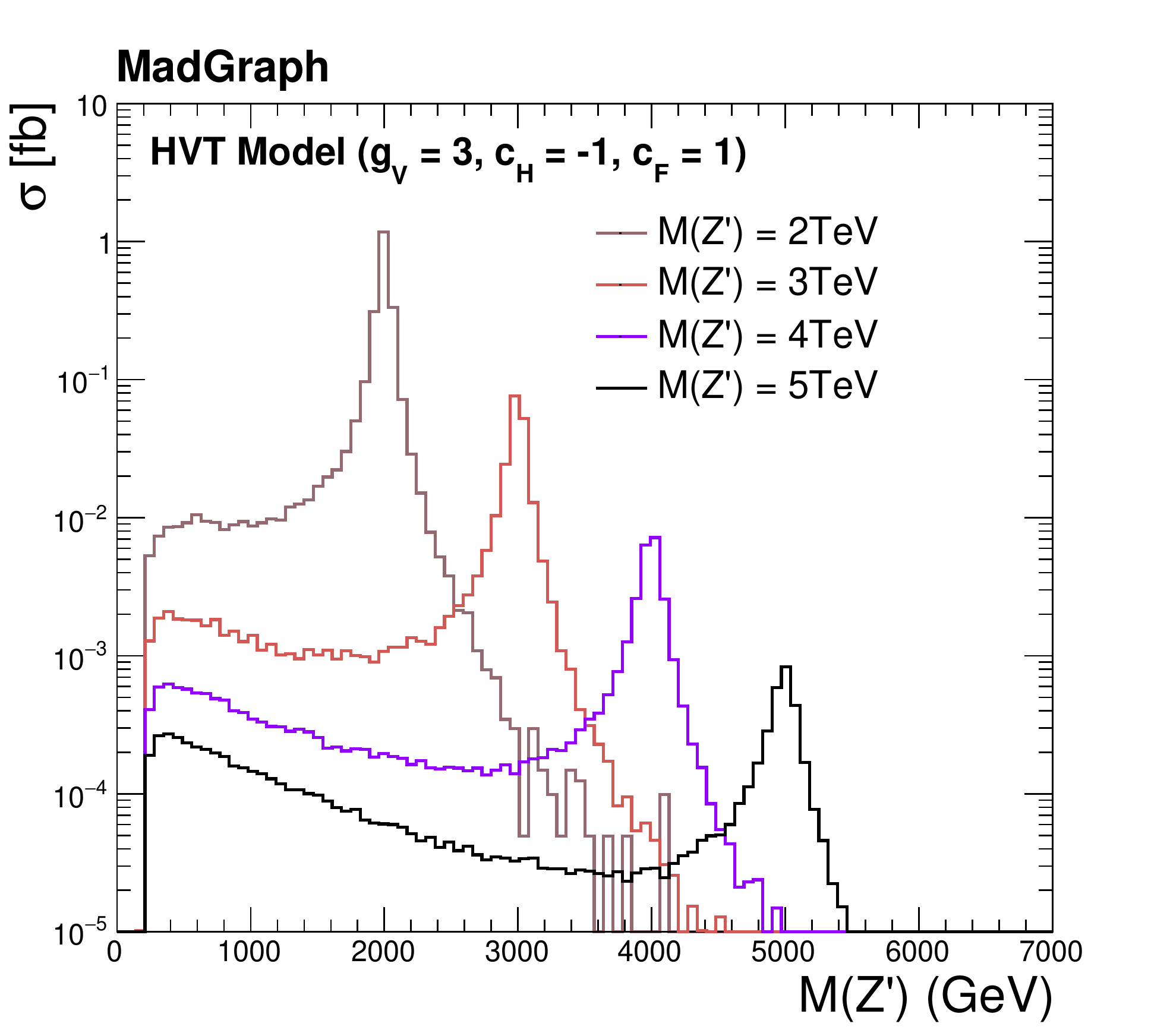}
}
\caption{
Leading order $Z^\prime$ resonance differential cross sections
$d\sigma/dM( Z^\prime )$ 
as predicted from the HVT model.
}
\label{Zprimegv3}
\end{center}
\vspace{-5mm}
\end{figure}
\begin{figure}[hbt!]
\begin{center}
\resizebox{0.5\textwidth}{!}{
\includegraphics{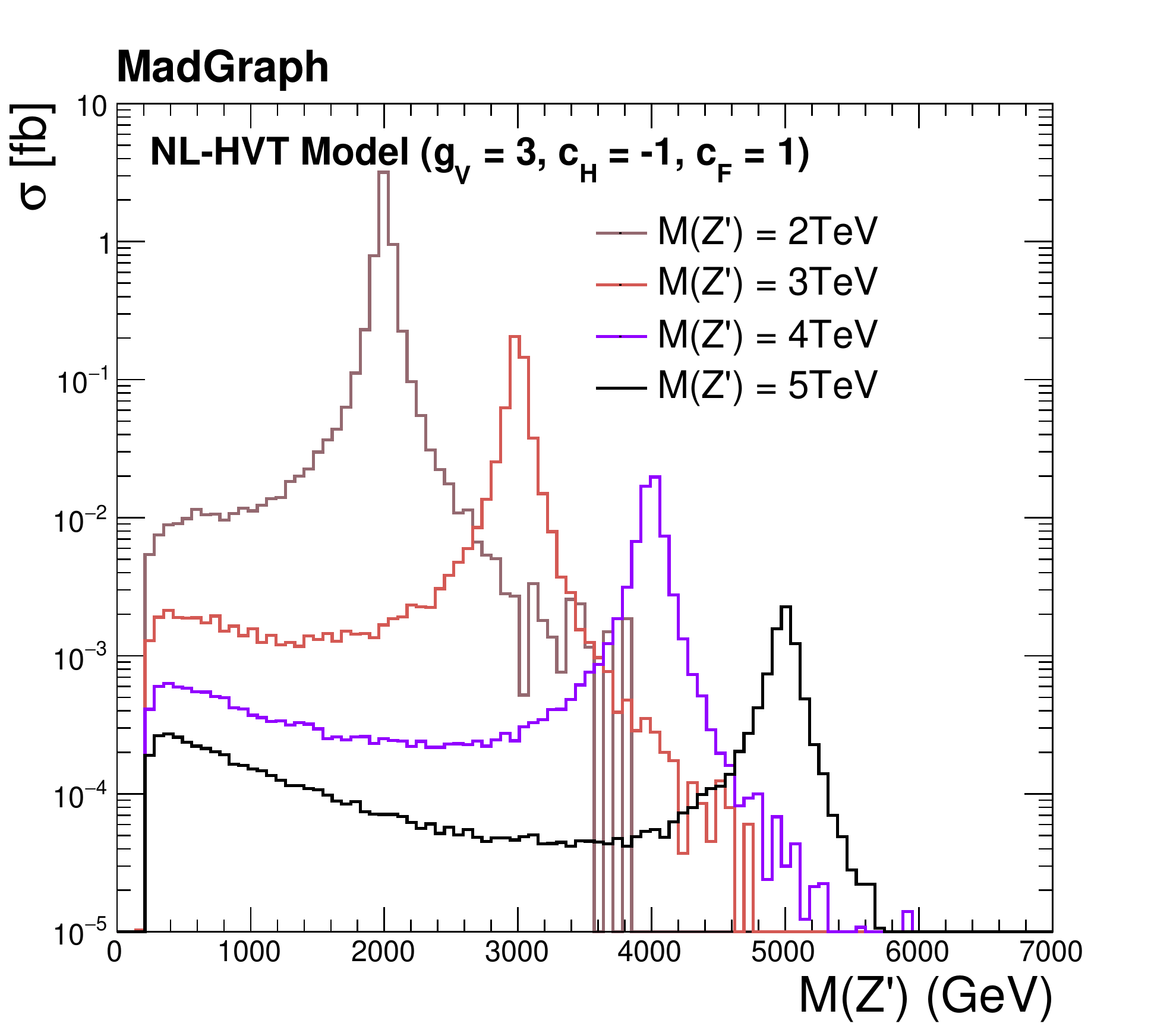}
}
\caption{
Leading order $Z^\prime$ resonance differential cross sections
$d\sigma/dM( Z^\prime )$ 
as predicted from the HVT model modified by the non-local form factor of Eq.~\ref{nlmod} with $a = 1$.
}
\label{Zprimegv3M}
\end{center}
\vspace{-5mm}
\end{figure}

In Fig.~\ref{NLxsection} sample $Z^\prime$ resonance DY cross sections as 
calculated from the HVT model and modified by the form factor 
of Eq.~\ref{nlmod} are shown.
The non-locality scale $\Lambda_{NL}$ has been 
set at the pole mass $M(Z^\prime)$.
Due to the presence of the factor $e^{a (s/\Lambda_{NL}^{2})}$, the
$Z^\prime$ lineshapes receive significant non-local modifications 
and are distorted. This is shown in Fig.~\ref{Zprimegv3} where the 
HVT model is used, and in Fig.~\ref{Zprimegv3M} where the non-local 
modification is applied (NL-HVT). 
The potential increase of the cross section
can lead to measureable anomalous Higgs boson production at very high $P_T$.
One can notice the presence of long low mass tails coming from the parton
density functions. 

%
%

The HVT predicted lineshape and cross-section
modification in the presence of a nearby non-locality scale
$\Lambda_{NL}$, leads to a rich phenomenology.
Depending on the number of resonances, their masses and widths,  
the impact on the $VH$ and $VV$ production at high momentum 
transfers could be significant. Thus, the interesting possibility 
that LHC phase-2 measurements could probe energy scales 
of order~10~TeV, is worth exploring.

\begin{figure}[hbt!]
\begin{center}
\resizebox{0.5\textwidth}{!}{
\includegraphics{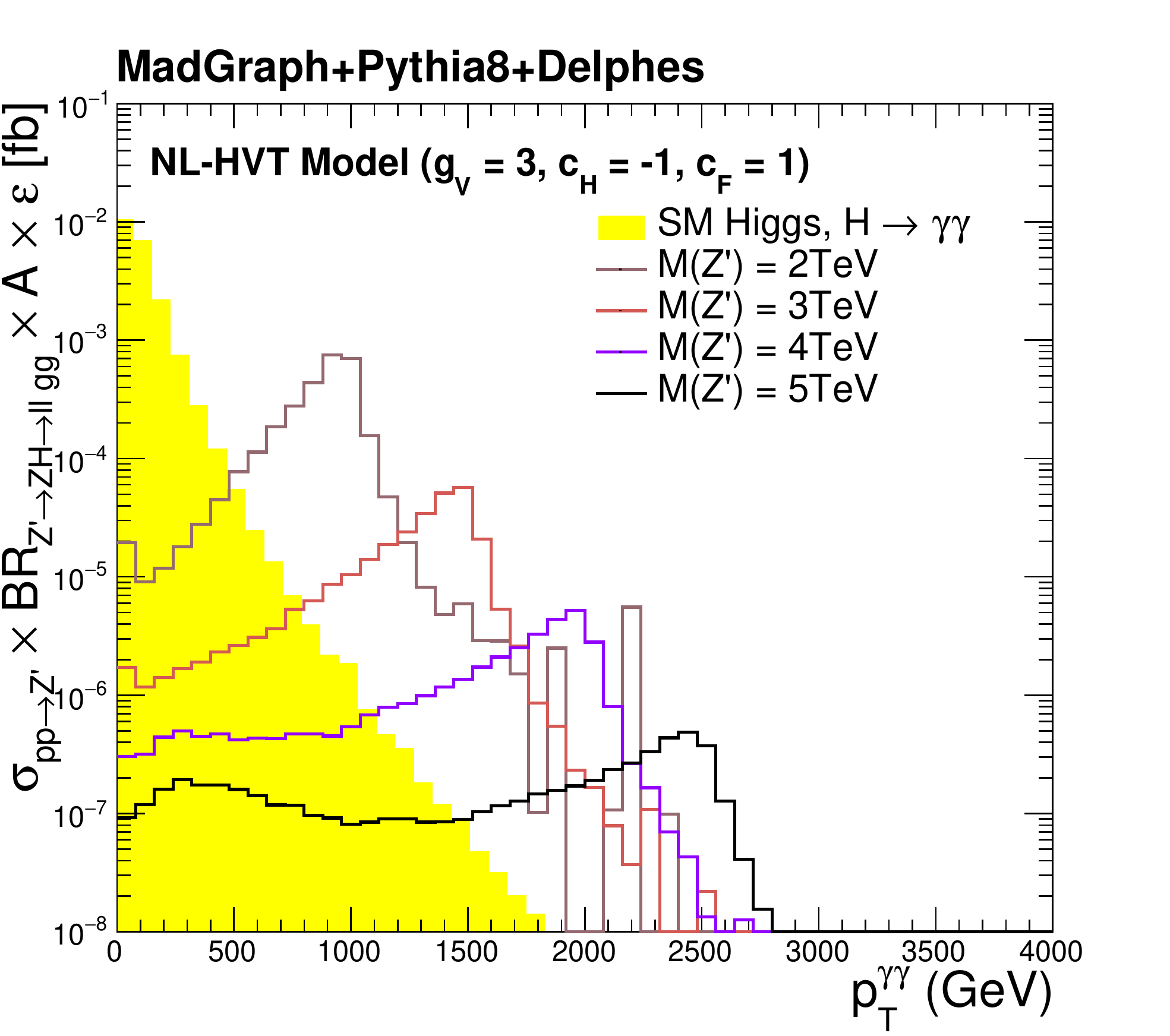}
}
\caption{
$Z^\prime\rightarrow ZH \rightarrow \ell\ell \gamma\gamma$
  Higgs yields with respect to diphoton transverse momentum
$dN/dP_T^{\gamma \gamma}$, in fb,
as predicted from the non-local form factor in Eq.~\ref{nlmod} with $a = 1$.
The expected SM Higgs yield for the same selection is overlayed in yellow.
}
\label{diphotonPt}
\end{center}
\vspace{-5mm}
\end{figure}

\begin{figure}[hbt!]
\begin{center}
\resizebox{0.5\textwidth}{!}{
\includegraphics{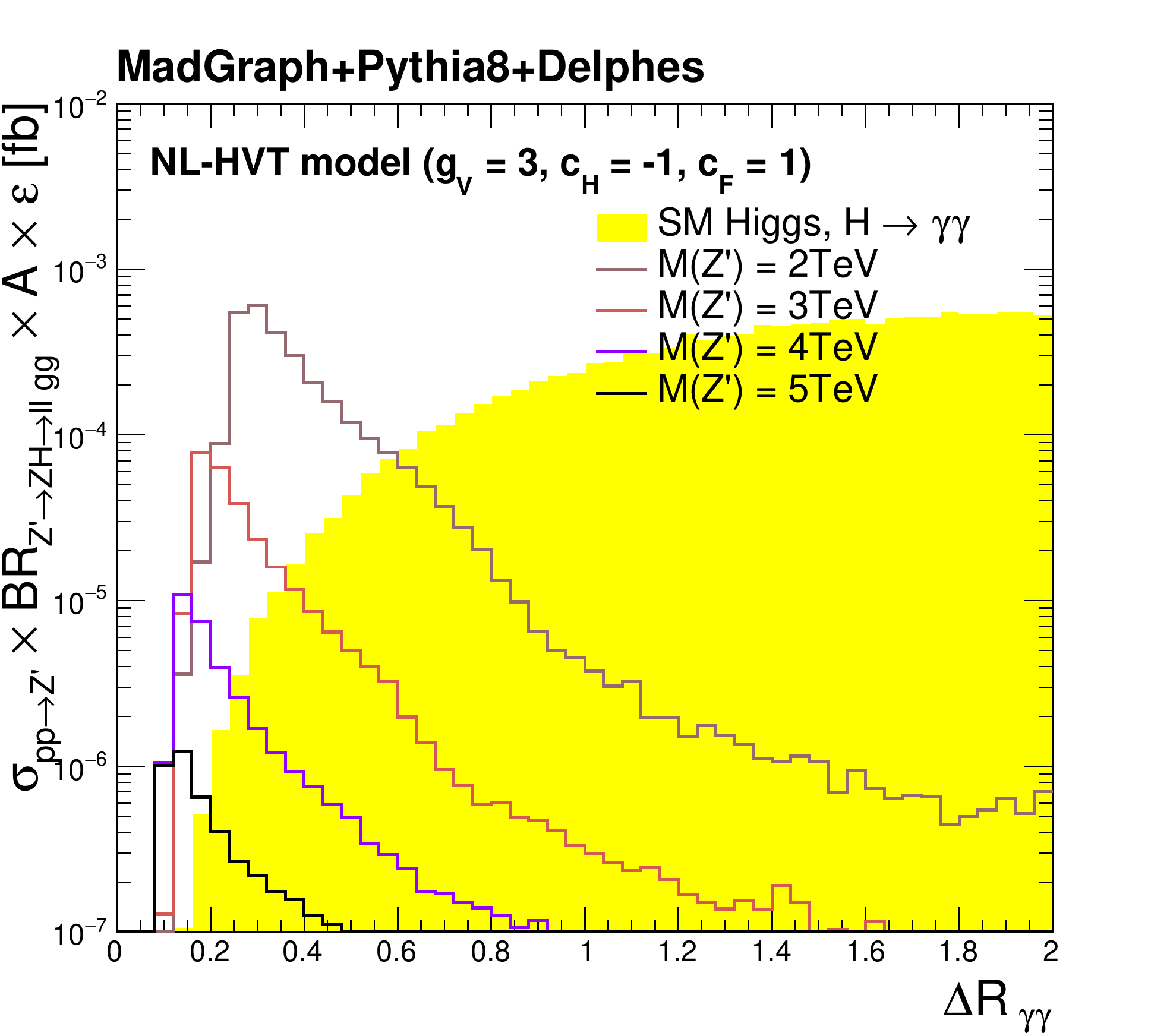}
}
\caption{
$Z^\prime\rightarrow ZH \rightarrow \ell\ell \gamma\gamma$
  Higgs yields with respect to $\Delta R_{\gamma \gamma}$,
$dN/d\Delta R_{\gamma \gamma}$, in fb,
as predicted from the non-local form factor in Eq.~\ref{nlmod}
with $a = 1$.
The expected SM Higgs yield for the same selection is overlayed in yellow.
}
\label{diphotonDR} 
\end{center}
\vspace{-5mm}
\end{figure}

\begin{figure}[hbt!]
\begin{center}
\resizebox{0.5\textwidth}{!}{
\includegraphics{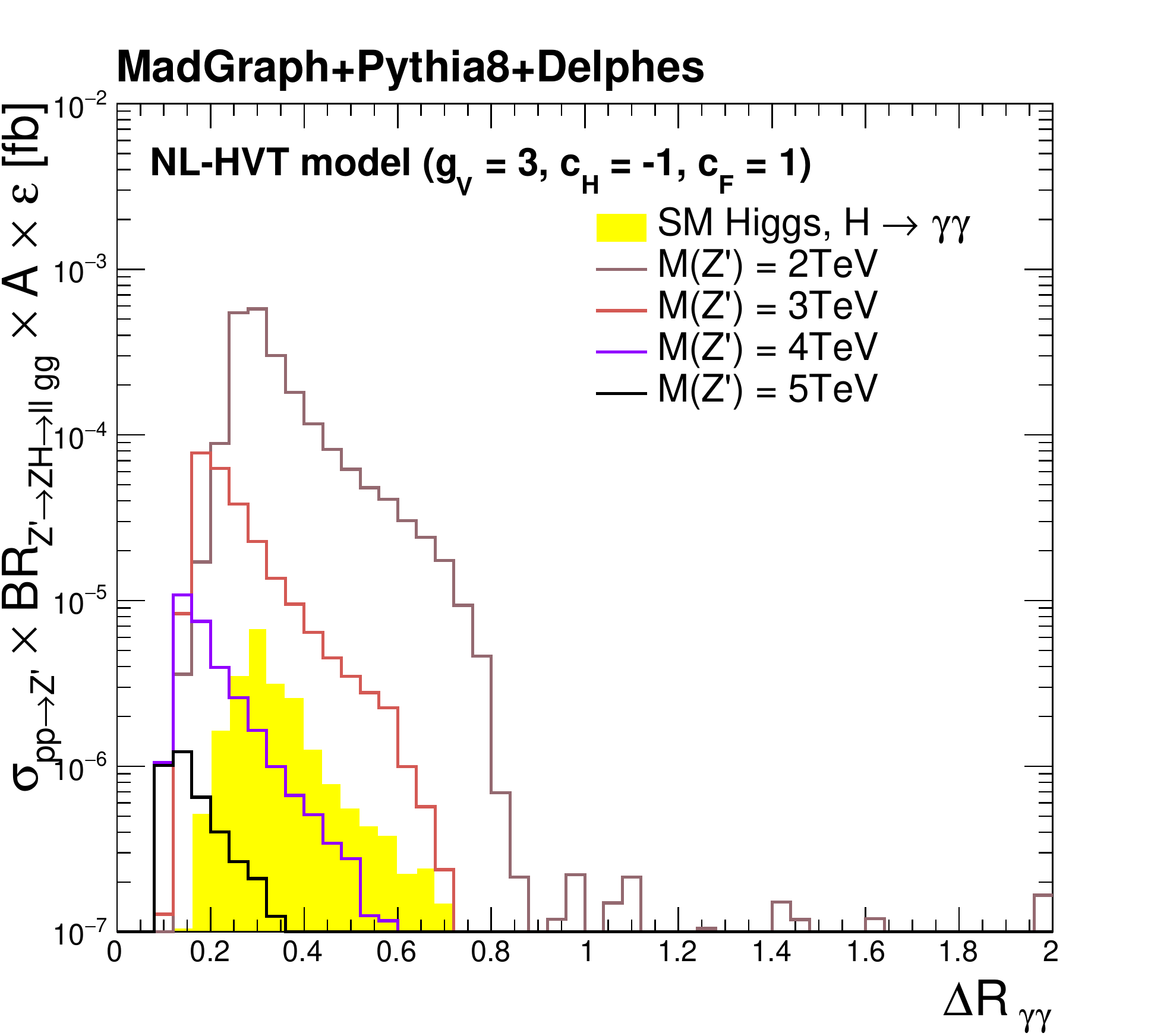}
}
\caption{
$Z^\prime\rightarrow ZH \rightarrow \ell\ell \gamma\gamma$
  Higgs yields with respect to $\Delta R_{\gamma \gamma}$,
$dN/d\Delta R_{\gamma \gamma}$, in fb, after a diphoton
  transverse momentum cut of $P_T>800$GeV,
as predicted from the non-local form factor in Eq.~\ref{nlmod} with $a = 1$.
The expected SM Higgs yield for the same selection is overlayed in yellow.
}
\label{DiphotonDRpt800} 
\end{center}
\vspace{-5mm}
\end{figure}

The diphoton system opening angle $\Delta R$ and transverse momentum 
$P_T$ are key variables sensitive to non-local effects.
In Fig.~\ref{diphotonPt} the Higgs yields as a function
of the diphoton $P_T$ are shown for
$Z^\prime$ masses of 2, 3, 4 and 5~TeV. 
The SM Higgs prediction is shown in yellow and 
demonstrates that at high transverse $P_T$, its contribution becomes
very small. The Higgs yields as a function of the diphoton $\Delta R$ 
are shown in
Figures~\ref{diphotonDR} and \ref{DiphotonDRpt800}  before any
transverse momentum selection and after a $P_T>800$~GeV cut, respectively.
Although the opening angle between
the two photons originating from the Higgs decay is typically small, 
we can see that even for higher $Z^\prime$ masses, 
the photons are still separated by $\Delta R > 0.15$. Such levels of 
angular separation allow the reconstruction of two close-by high-$P_T$
photon clusters in the electromagnetic calorimeter. The presence 
of a pair of high-$P_T$ photons in the detector with an invariant mass close to 
125~GeV, is a signal for new physics.
%
%
%

%

%
%
\section{Sensitivity of a search at LHC}
\label{analysis}
The observable that can serve as an early signal of non-locality at the 
LHC is the anomalous production of Higgs bosons at very high $P_T$.
Although the $H\rightarrow b\bar{b}$ is a potentially powerful decay channel 
for measuring such anomalous production,  
for Higgs $P_T > 1$~TeV the sensitivity drops due to the merging of the two $b$ 
jets in a single jet. The efficiency of identifying small opening
angle $b\bar{b}$ pairs falling into a single fat jet drops
fast with the transverse $P_T$. In addition, for a single $b\bar{b}$ jet, 
the QCD background is quite significant. 

In the case of the diphoton channel, although the branching ratio 
$H\rightarrow\gamma \gamma$ is 213 times lower than the
$H\rightarrow b\bar{b}$, the photon pairs are potentially 
separable even at very high Higgs $P_T$ while the diphoton
identification efficiency is
very high. The expected small continuum background from
diphoton pairs at 125~GeV,
the very small SM Higgs yields, and the excellent diphoton mass resolution
lead to a competitive sensitivity when compared to $H\rightarrow b\bar{b}$.  
We argue that the observation of even a single high $P_T$ diphoton 
pair with invariant mass close to the Higgs mass, would be
statistically significant.
Considering that the anomalous Higgs production could be produced by 
a broad heavy resonance or a continuum, one can proceed with
the reconstruction of its full
invariant mass distribution.
However, this would require a large number of signal events. 
The search we propose has a SM Higgs background expectation that is 
at least one order of magnitude lower than the signal,
aiming in observing at the LHC phase 2 the first signs of new 
physics by looking for 
events that have very small likelihood to be SM-induced. In this
work we propose a search for boosted Higgs bosons decaying into 
two photons in association with a weak gauge boson.

To evaluate the expected signal yields for a broad
range of masses and widths, we rely on Monte Carlo
samples generated with MadGraph5, \cite{madgraph5}, interfaced to
PYTHIA \cite{Pythia8:2008}. To simulate the response of an
LHC-like experiment, realistic resolution and reconstruction
efficiencies for electrons, muons, photons and jets were applied with
the Delphes framework \cite{delphes}. 
The specific cut-based event selection was based on a simplification
of the cuts performed by LHC experiments on SM $VH$ searches with
$H\rightarrow \gamma\gamma$ \cite{CMS-Hgg2021}:
\begin{itemize}
\item
We require at least two photon candidates with the two leading candidates
having $P_T^{\gamma 1}>m_{\gamma\gamma}/3$ and $P_T^{\gamma 2}>m_{\gamma\gamma}/4$.
\item
The diphoton invariant mass must be in the range
$120<m_{\gamma\gamma}<130$~GeV, called the signal region.
\item
Two opposite sign leptons with $P_T^{l1}>30$~GeV and $P_T^{l2}>30$~GeV are required for the $ZH$ case, and a single 
lepton with $P_T^{l1}>30$~GeV for the $WH$ case.
\item
For the $ZH$ mode, the dilepton invariant mass is restricted to be close
to the $Z$ pole mass: $80<m_{\ell}<110$~GeV. 
\end{itemize}

The difference between this selection and the 
SM diphoton reconstruction is that in our case
the two photons have a small opening angle, so any photon isolation
cuts have to be relaxed in order to preserve high reconstruction
efficiency.
An important additional requirement is a $W^\prime/Z^\prime$ mass-dependent cut 
on the transverse momentum of the $\gamma\gamma$ system, 
$P_T>M_{V^\prime}/3$, that selects highly boosted SM Higgs candidates. 
For such high diphoton $P_T$, both photons and leptons are also of very
high transverse momentum and the event topology is really spectacular:
two relatively small opening angle high $P_T$ photons and two (one) 
high $P_T$ leptons for the $Z^\prime\rightarrow ZH$
($W^\prime\rightarrow WH$) channel. For such high $P_T$ objects the
single-object reconstruction efficiency is above 95\% and most of 
the efficiency losses come from the event selection.
The expected signal and background yields in units of fb are 
shown in Table \ref{yields1}.
%
\begin{table*}[t]
\small
  \begin{center}
    \begin{tabular}{lcccc}\hline
 Process           
& $\sigma\times BR$ & A$\times \epsilon$ & Yield ($P_T^{\gamma \gamma} > M_{V^{\prime}}/3$)\\
 $qq\rightarrow V^\prime\rightarrow \gamma\gamma$ 
&  [fb]             & [\%]               & [fb]  & $M_{V^{\prime}}$ [TeV]\\
\hline
$Z^{\prime}\rightarrow ZH\rightarrow \ell\ell \gamma\gamma$ (2~TeV) 
& $1.90\times 10^{-3}$ & 40.8 & $6.8\times 10^{-4}$ & 2\\
$Z^{\prime}\rightarrow ZH\rightarrow \ell\ell \gamma\gamma$ (3~TeV) 
& $1.97\times 10^{-4}$ & 34.4 & $5.3\times 10^{-5}$ & 3\\
$Z^{\prime}\rightarrow ZH\rightarrow \ell\ell \gamma\gamma$ (4~TeV) 
& $2.99\times 10^{-5}$ & 29.9 & $5.1\times 10^{-6}$ & 4\\
$Z^{\prime}\rightarrow ZH\rightarrow \ell\ell \gamma\gamma$ (5~TeV) 
& $6.36\times 10^{-6}$ & 29.4 & $3.8\times 10^{-7}$ & 5\\
$W^{\prime}\rightarrow WH\rightarrow \ell\nu \gamma\gamma $ (2~TeV)
& $1.26\times 10^{-2}$ & 54.1 & $6.0\times 10^{-3}$ & 2\\
$W^{\prime}\rightarrow WH\rightarrow \ell\nu \gamma\gamma $ (3~TeV)
& $1.33\times 10^{-3}$ & 46.7 & $4.8\times 10^{-4}$ & 3\\
$W^{\prime}\rightarrow WH\rightarrow \ell\nu \gamma\gamma $ (4~TeV)
& $1.99\times 10^{-4}$ & 42.3 & $4.8\times 10^{-5}$ & 4\\
$W^{\prime}\rightarrow WH\rightarrow \ell\nu \gamma\gamma $ (5~TeV)
& $4.20\times 10^{-5}$ & 43.8 & $5.4\times 10^{-6}$ & 5\\
\hline
SM $ZH\rightarrow \ell\ell \gamma\gamma $ 
& $0.20$ & 13.3 & $3.8\times 10^{-5}$ & 2\\
SM $ZH\rightarrow \ell\ell \gamma\gamma $ 
& $0.20$ & 13.3 & $3.4\times 10^{-6}$ & 3\\
SM $ZH\rightarrow \ell\ell \gamma\gamma $ 
& $0.20$ & 13.3 & $5.1\times 10^{-7}$ & 4\\
SM $ZH\rightarrow \ell\ell \gamma\gamma $ 
& $0.20$ & 13.3 & $7.0\times 10^{-8}$ & 5\\
SM $WH\rightarrow \ell\nu \gamma\gamma $ 
& $1.01$ & 35.2 & $2.7\times 10^{-4}$ & 2\\
SM $WH\rightarrow \ell\nu \gamma\gamma $ 
& $1.01$ & 35.2 & $2.5\times 10^{-5}$ & 3\\
SM $WH\rightarrow \ell\nu \gamma\gamma $ 
& $1.01$ & 35.2 & $3.3\times 10^{-6}$ & 4\\
SM $WH\rightarrow \ell\nu \gamma\gamma $ 
& $1.01$ & 35.2 & $3.7\times 10^{-7}$ & 5\\
\hline
SM Continuum $\ell\ell \gamma\gamma $ 
& $638.2$ & 0.21 & $2.7\times 10^{-4}$ & 2\\
SM Continuum $\ell\ell \gamma\gamma $ 
& $638.2$ & 0.21 & $1.6\times 10^{-5}$ & 3\\
SM Continuum $\ell\ell \gamma\gamma $ 
& $638.2$ & 0.21 & $2.8\times 10^{-6}$ & 4\\
SM Continuum $\ell\ell \gamma\gamma $ 
& $638.2$ & 0.21 & $5.6\times 10^{-8}$ & 5\\
SM Continuum $\ell\nu \gamma\gamma $ 
& 654.4 & 2.9 & $5.2\times 10^{-4}$ & 2\\
SM Continuum $\ell\nu \gamma\gamma $ 
& 654.4 & 2.9 & $4.1\times 10^{-5}$ & 3\\
SM Continuum $\ell\nu \gamma\gamma $ 
& 654.4 & 2.9 & $3.4\times 10^{-6}$ & 4\\
SM Continuum $\ell\nu \gamma\gamma $ 
& 654.4 & 2.9 & $1.6\times 10^{-7}$ & 5\\

    \hline
    \end{tabular}
    \caption{Non-local anomalous Higgs production yields for $a=1$ in fb, 
compared to SM Higgs and continuum SM background yields for
searches targeting resonance masses of 2, 3, 4 and 5~TeV.}
    \label{yields1}
  \end{center}
\end{table*}
%
%
In the calculations shown in Table~\ref{yields1},
the continuous background $\ell\ell\gamma\gamma$ expectations 
are shown for a diphoton transverse momentum cut, 
$P_T^{\gamma\gamma} > M_{V^{\prime}}/3$.
A potentially significant background is the
$\ell\ell\gamma$ production with an additional fake photon, which 
is also included in the analysis.
Fake rejection can only be estimated with full detector simulation 
and data, thus the $\ell\ell\gamma$ prediction from the fast simulation is not 
realistic and results to an overestimation of this background.
The final study and further cut optimizations are left to the 
LHC experiment groups.
 
Based on the results of Table~\ref{yields1}, we
conclude that for very high diphoton $P_T$, 
the SM Higgs production and the continuum
background dominated by $Z/W + \gamma\gamma$, are the dominant
backgrounds.
Their contribution in the signal region of $120-130$~GeV is rather small. 
The proposed analysis starts becoming sensitive for yields in 
the range of a few $10^{-3}$~fb. This means that with an integrated luminosity
of 300~fb$^{-1}$ from 
the LHC phase~1, we should start observing very high $P_T$ Higgs candidates
recoiling off a dilepton or a single lepton. The High Luminosity 
LHC phase 2, is definitely in a better position to probe the NL-SM phase space, 
since we expect an integrated luminosity of 3000~fb$^{-1}$.
However, signal yields are still in the
sub-femtobarn level, thus requiring significant amounts of
luminosity to observe a single event in the signal region.

\begin{figure}[hbt!]
\begin{center}
\resizebox{0.5\textwidth}{!}{
 \includegraphics{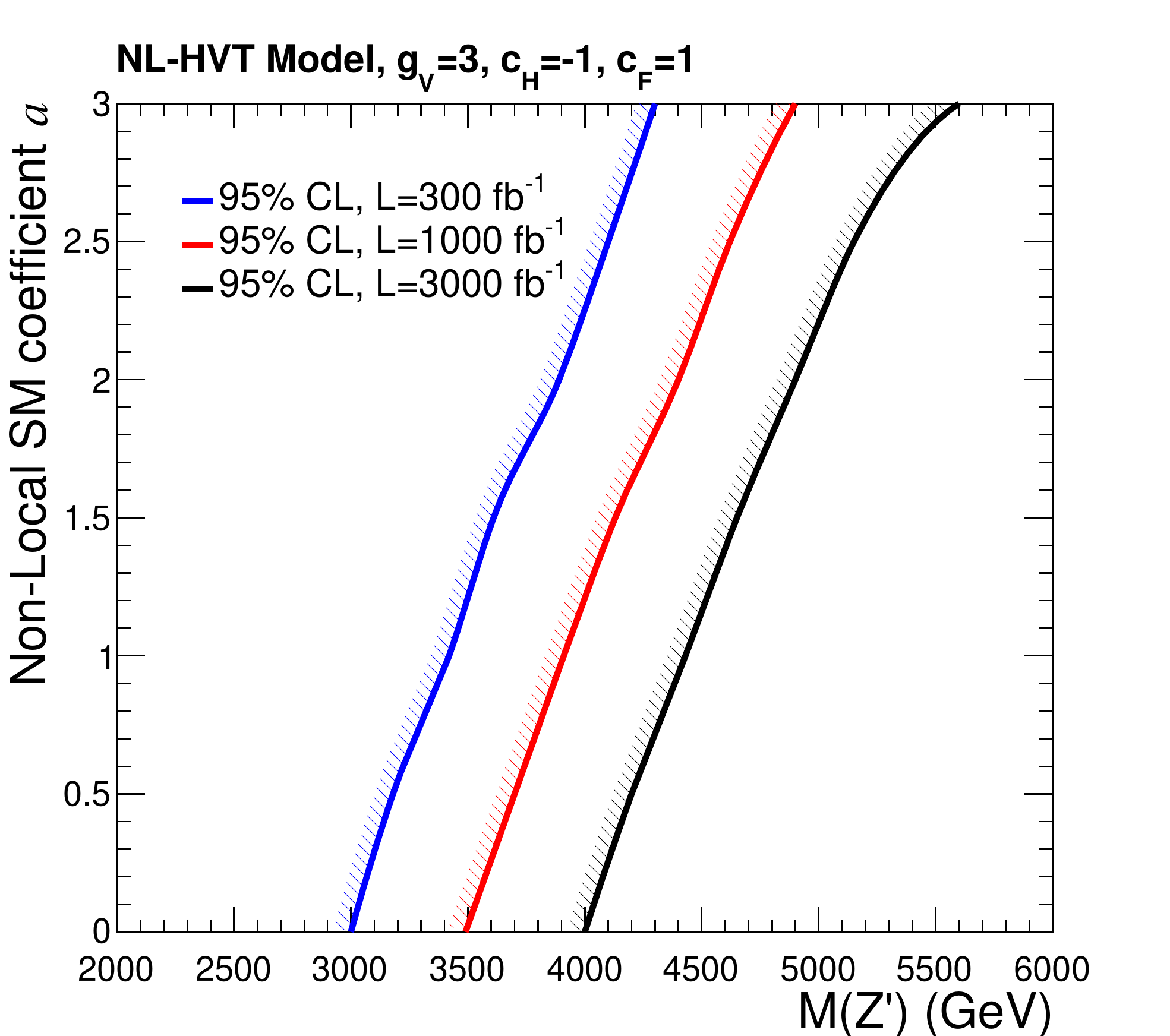}
}
\caption{
95\% Confidence Level limits in non-locality scale $\Lambda_{NL}$ as a
function of the parameter $a$ and the integrated luminosity, 
based on measurements of anomalous Higgs production from a $Z^\prime $ resonance.
Non-local modifications are obtained using the form factor $e^{a (s/\Lambda_{NL}^2)}$.
}
\label{Sensit1} 
\end{center}
\vspace{-5mm}
\end{figure}

\begin{figure}[hbt!]
\begin{center}
\resizebox{0.5\textwidth}{!}{
 \includegraphics{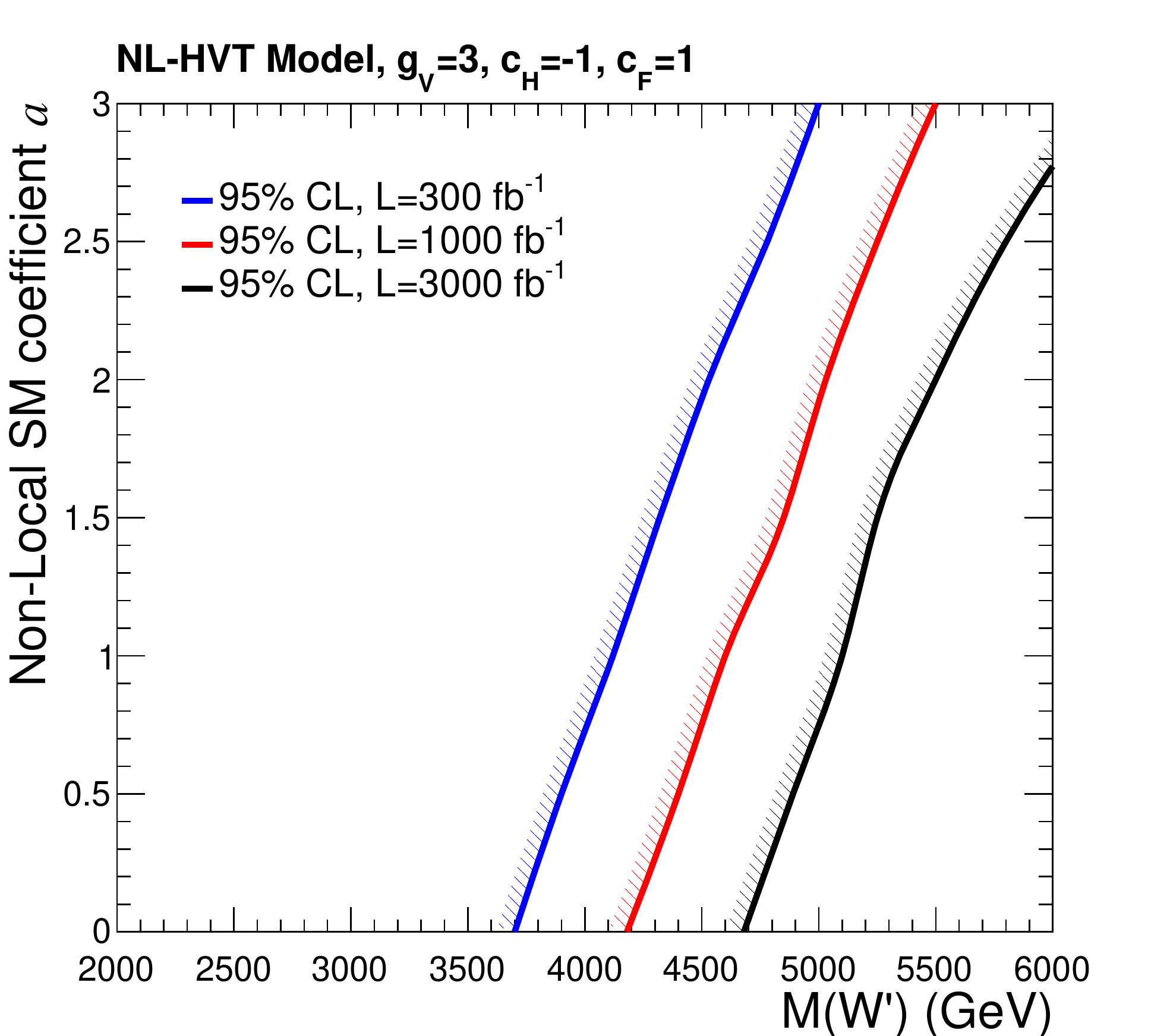}
}
\caption{
95\% Confidence Level limits in non-locality scale $\Lambda_{NL}$ as a
function of the parameter $a$ and the integrated luminosity, 
based on measurements of anomalous Higgs production from a $W^\prime $ resonance.
Non-local modifications are obtained using the form factor $e^{a (s/\Lambda_{NL}^2)}$.
}
\label{Sensit2} 
\end{center}
\vspace{-5mm}
\end{figure}

The sensitivity of a search for non-local effects using high $P_T$ Higgs as
a probe, is presented as a function of the integrated luminosity and the scale of new
physics in Figures~\ref{Sensit1} and \ref{Sensit2} \cite{asymp}. To fully scan the parameter space, we
allow for the parameter $a$ in Eq.~\ref{nlmod} to vary. 
The case $a=0$ corresponds to the
pure HVT model, i.e. absense of non-local modifications. 
From these results, one can see that the boosted Higgs $VH$ leptonic mode 
can prove a powerful probe for anomalous high $P_T$ Higgs production. 
For the high luminosities expected in the phase 2 of LHC the
sensitivity of the search is expected to reach 4-5~TeV 
for $a\simeq 1$.

%
%
%
%
%
%
%
%
%

\section{Summary and Conclusions}
In this work we explored the potential of observing early signs of 
non-local SM effects at the LHC. We have assumed that the non-locality 
scale is of order of a few TeV, and that at this scale heavy vector
bosons are present. The heavy vector-boson cross section is
modified by the presence of a non-local scale, leading to significant 
modification of their lineshape. Assuming the general case of
resonances with finite width we calculate their cross sections using
a simple model (HVT) that is modified by $\sim e^{a (s/\Lambda_{NL}^2)}$
factors. Although the
non-locality scale does not have to match the vector boson pole mass
$M_{V^{\prime}}$, observing that in the case of QCD the $\rho$ mass is
very close to the QCD transition scale, we have assumed
$M_{V^{\prime}}=\Lambda_{NP}$.

Possible modifications of dilepton invariant mass continuum at high $Q^2$
already place constraints on non-local QFT, however the modifications 
of new heavy resonance lineshapes or their continuum could have more
dramatic consequenses due to their proximity to the new physics scale.
The first signs of the presence of non-locality could be through the $VH$ 
channel with the $W/Z$ as well as the Higgs boson displaying TeV-level 
transverse momentum. This channel leads to ve\-ry clean sear\-ches 
with small expected SM backgrounds. Although, in this work we focused on the 
single lepton and dilepton channels, the $Z\rightarrow \nu\nu$ channel
and the $Z/W$ hadronic channels may significantly improve the search.
In particular the $Z\rightarrow \nu\nu$ channel can offer a
spectacular signature of a boosted Higgs recoiling off a TeV-scale
missing $E_T$.

Based on our results and due to the limitation of the LHC 
centre-of-mass 
energy, we only expect to probe non-local effects if
the relevant scale is not much higher than 5~TeV. 
At the LHC, a programme aimed at
measuring Higgs production at very high $P_T$ has 
already started, and the upcoming data from 
run~3 and LHC phase~2 are
expected to probe locality at new record small distance scales.


\section*{Acknowledgements}
This work was supported by the Taiwanese Ministry of Science and Technology 
grant 109-2112-M-002-014-MY3.



\end{document}